# Carnot heat engine efficiency with a paramagnetic gas

Manuel Malaver[1,2,3]

1 Bijective Physics Institute, Bijective Physics Group, Idrija, Slovenia
2 Entro-π Fluid Research Group.
3 Maritime University of the Caribbean, Catia la Mar, Venezuela.

*Corresponding author (Email: mmf.umc@gmail.com)

**Abstract –** Considering ideal paramagnetic medium, in this paper we deduced an expression for the thermal efficiency of Carnot heat engine with a paramagnetic gas as working substance. We found that the efficiency depends on the limits of maximum and minimal temperature imposed on the Carnot cycle, as in the ideal gas, photon gas and variable Chaplygin gas.

## 1. Introduction

The magnetism is a type of associated physical phenomena with the orbital and spin motions of electrons and the interaction of electrons with each other (Ahmed et al., 2019). The electric currents and the atomic magnetic moment of elementary particles generate magnetic fields that can act on other currents and magnetic materials. The most common effect occurs in ferromagnetic solids which is possible when atoms are rearrange in such a way that magnetic atomic moments can interact to align parallel to each other (Eisberg and Resnick, 1991; Chikazumi, 2009). In quantum mechanics, the ferromagnetism can be described as a parallel alignment of magnetic moments what depends of the interaction between neighbouring moments (Feynman et al., 1963).

Although ferromagnetism is the cause of many frequently observed magnetism effects, not all materials respond equally to magnetic fields because the interaction between magnetic atomic moments is very weak while in other materials this interaction is stronger (Ahmed et al., 2019). In the atoms and molecules of paramagnetic materials there are unpaired electrons which freely align their magnetic moment in any direction and when an external field is applied these magnetic moments align in the same direction of applied field making it more intense (Eisberg and Resnick, 1991). Hydrogen, lithium and oxygen are examples of paramagnetic substances.

The paramagnetic property of oxygen is important in many processes (Miessler and Tarr, 2010). As the paramagnetism is strongly temperature-dependent, cold oxygen molecules are attracted by a strong magnetic field and when they are heated leave the magnetic field, this gives rise to a current which generates a measure of the oxygen content.

Diamagnetism is an intrinsic property of all materials and is the tendency of materials to be repelled by an applied magnetic field because of the presence no unpaired electrons so that the atomic magnetic moments not produce any effect (Jackson, 2014). The other forms of magnetism as paramagnetism or ferromagnetism are much stronger in a material and the diamagnetic contribution is very negligible (Ahmed et al., 2019). Examples of diamagnetic substances are Helium, bismuth and water.

A paramagnetic gas can be treated as an ideal monoatomic gas (García-Colín Scherer, 1972). An ideal gas is a gas composed of a group of randomly moving, non-interacting point particles. The ideal gas approximation is useful because it obeys the gases laws and represent the vapor phases of fluids at high temperatures for which the heat engines is constructed (Lee, 2001). Any device for converting heat into work in a cyclic process can be called a heat engine or thermal machine and must operate in the presence of two different temperatures (Dickerson, 1969). In a steam engine, for example, the high temperature is the temperature of the steam and the low temperature is the condensed cold. A heat engine that can work with an ideal gas as working substance is the Carnot cycle. For the ideal gas, Carnot cycle will be composed by two isothermal curves and adiabatic which will come given by the conditions $PV = const$ and $PV^\gamma = const$, respectively (Dickerson, 1969; Nash, 1970) where $\gamma$ is an adiabatic exponent.

One of the great virtues of the Carnot cycle is its potential applicability to any working substance (Nash, 1970). In agreement with Leff (2002) and Lee (2001) the Carnot cycle for a photon gas provides a very useful tool to illustrate the thermodynamics laws and it is possible to use for introducing the concepts of creation and annihilation of photons in an introductory course of physics. Bender et al. (2000), showed that the efficiency of a quantum Carnot cycle is the same as that of a classical Carnot cycle, with the identification of the expectation value of the Hamiltonian as the temperature of the system. Unlike the ideal gas, the pressure for a photon gas is a function only of the temperature and the internal energy function is dependent of volume (Leff, 2002). Malaver (2015,2018) found that thermodynamic efficiency of Carnot cycle for a variable Chaplygin gas depend only on the limits of maximum and minimal temperature imposed to cycle as in case of the ideal gas an photon gas.

In this paper, an expression for the efficiency of Carnot cycle with a paramagnetic gas as working substance is deduced from the equation of state for an ideal gas. We have found that the efficiency of Carnot cycle in a paramagnetic gas will depend on the limits of maximum and minimal temperature imposed on the cycle. The article is organized as follows: in Section 2, the physical properties of Carnot heat engine are studied; in Section 3, we show the deduction for the thermal efficiency of Carnot cycle for the ideal gas; in Section 4, we obtain an expression for the efficiency of Carnot engine with a paramagnetic gas; in Section 5, we conclude.

## 2. Carnot heat engine

In the process of operation of a heat engine between two different temperatures, some heat is always transferred on the outside (Dickerson, 1969). If an amount of heat $Q_H$ is absorbed at the high temperature the work is done on the surroundings, and if a quantity of heat $Q_L$ is lost at the lower temperature, then of the first law of thermodynamics for the entire cyclic process $Q_H + Q_L + W_{net} = \Delta U = 0$ where $\Delta U$ is the variation of the internal energy in the process and $W_{net}$ is the work in the cycle (Dickerson,1969; Nash,1970). The efficiency of a heat engine is commonly defined as the ratio of the work obtained from the system to the heat taken from the hot reservoir

$$\eta = \frac{-W_{net}}{Q_H} = \frac{Q_H + Q_L}{Q_H} = 1 + \frac{Q_L}{Q_H}$$

(1)

A particularly simple heat engine cycle to handle mathematically is the Carnot cycle (Dickerson,1969 ; Wark and Richards,2001). In the figure 1, two temperatures are included, $T_H$ and $T_L$. The first step in a Carnot cycle is a reversible isothermal expansion a $T_H$ or from point A to point B in figure 1. This expansion could be achieved by expanding the gas in contact with a large heat reservoir at $T_H$. A certain amount of work will be done on the surroundings which implies an absorption of heat .

The second step is an reversible adiabatic expansion from the state at point B at point C. Under these conditions $Q_{BC} = 0$ and $\Delta U_{BC} = W_{BC}$ and the internal energy change is the same as the work done on the working substance. Since work is done on the surroundings, $W_{BC}$ is negative and the internal energy must fall.

The third step, the reversible isothermal compression, is continued just to the point C where a final adiabatic compression will bring the gas back to its starting conditions at point A on the $PV$ plot of the figure 1. Work $W_{CD}$ is done on the gas, and an amount of heat $Q_L$ is lost from the gas which compensates for this work exactly in an ideal gas and approximately in a real gas.

In agreement with Dickerson (1969) the final adiabatic compression to the starting point occurs with work $W_{DA}$ done on the gas and an increase in the internal energy. For the entire cycle, $W_{net} = W_{AB} + W_{BC} + W_{CD} + W_{DA}$ and $Q_{neto} = Q_H + Q_L$. The total sum of heat and works is zero since the initial and final states are identical

$$Q_H + Q_L + W_{AB} + W_{BC} + W_{CD} + W_{DA} = Q_{net} + W_{net} = \Delta U = 0 \qquad (2)$$

The efficiency of the entire cycle in converting heat to work is

$$\eta = \frac{-W_{net}}{Q_H} = \frac{Q}{Q_H} = \frac{Q_H + Q_L}{Q_H} = 1 + \frac{Q_L}{Q_H} \qquad (3)$$

Since $Q_H$ and $Q_L$ have opposite signs, the efficiency is less than 1 and is the greatest possible efficiency.

## 3. Carnot cycle in an Ideal Gas

In this work, we have used the convention of Wark and Richards (2001) that defines the work during a reversible process as

$$W = -\int PdV \tag{4}$$

Following Dickerson (1969) and Nash (1970), in Fig. 1 we show the Carnot cycle for an ideal gas. In the first step of A to B, that is the isothermal expansion, there is no change in the internal energy $\Delta U$ in an ideal gas. This implies that

$$-W_{AB} = Q_{AB} = RT_H \ln \frac{V_B}{V_A} \tag{5}$$

$Q_{AB}$ is the absorbed heat in the first step, $T_H$ is the high temperature and R is the universal gas constant.

The second step of B to C is an adiabatic expansion. In this expansion $Q_{BC} = 0$ and the change in internal energy is equal to the work done

$$\Delta U_{BC} = W_{BC} = C_V(T_L - T_H) \tag{6}$$

where $C_V$ is the thermal capacity at constant volume and $T_L$ is the low temperature.

In the isothermal compression of C to D, the internal energy change is again zero and we obtain

$$-W_{CD} = Q_{CD} = RT_L \ln \frac{V_D}{V_C} \tag{7}$$

In the final adiabatic compression of D to A $Q_{DA} = 0$ and

$$\Delta U_{DA} = W_{DA} = C_V(T_H - T_L) \tag{8}$$

In a Carnot cycle for an ideal gas the net work done in the two adiabatic processes is zero and the adiabatic steps are related by the equation

$$\frac{V_B}{V_C} = \frac{V_A}{V_D} = \left(\frac{T_L}{T_H}\right)^{C_V/R} \tag{9}$$

The net work of the four steps is $W_{net} = W_{AB} + W_{CD}$. Substituting (9) into eqs. (5) and (7), we obtain

$$W_{net} = -R(T_H - T_L)\ln \frac{V_A}{V_B} \tag{10}$$

The heat absorbed at the high temperature is $Q_{AB}$ and the thermal efficiency of the entire cycle is

$$\eta = \frac{-W_{net}}{Q_{AB}} = 1 - \frac{T_L}{T_H} \qquad (11)$$

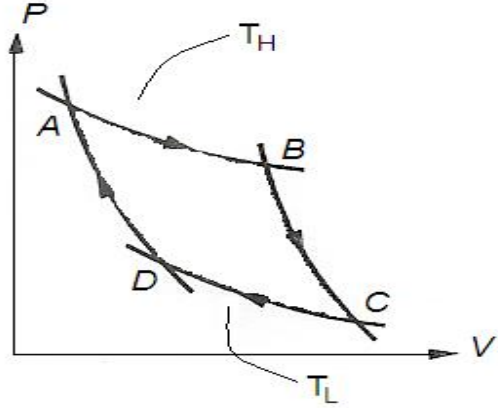

**Fig. 1.** Carnot cycle for an ideal gas

## 4. Thermal efficiency in an ideal paramagnetic gas

According García-Colín Scherer (1972) for an ideal paramagnetic gas, the first law of thermodynamics can be expressed as

$$dU = TdS - pdV + \mu_0 HdM \qquad (12)$$

$\mu_0$ is the vacuum permeability. In this paper were considered the equations of state for the ideal gas and the ideal paramagnetic medium

$$PV = RT \qquad (13)$$

$$M = C\frac{H}{T} \qquad (14)$$

Where $M$ is the magnetization, $T$ is the temperature, $H$ is the intensity of magnetic field applied and $C$ is the Curie's constant.

In a reversible adiabatic process $dS=0$ and for the equation (12) we have

$$dU = -PdV + \mu_0 HdM \tag{15}$$

Following Dickerson (1969), the internal energy $U(T,V)$ is given by

$$dU = \left(\frac{\partial U}{\partial V}\right)_T dV + \left(\frac{\partial U}{\partial T}\right)_V dT \tag{16}$$

But for the ideal gas $\left(\frac{\partial U}{\partial V}\right)_T = 0$ and $dU = \left(\frac{\partial U}{\partial T}\right)_V dT$ with $C_V = \left(\frac{\partial U}{\partial T}\right)_V$ and for the equation (15) can be written

$$C_V dT = -\frac{RT}{V}dV + \mu_0 HdM \tag{17}$$

For the ideal monoatomic gas $C_V = \frac{3}{2}R$ and substituting (14) in (17)

$$\frac{3}{2}RdT = -\frac{RT}{V}dV + \mu_0 \frac{MT}{C}dM \tag{18}$$

Rearranging eq.(18) we obtain

$$\frac{3}{2}R\frac{dT}{T} = -\frac{R}{V}dV + \mu_0 \frac{M}{C}dM \tag{19}$$

Integrating

$$\frac{3}{2}R\ln\frac{T_2}{T_1} = -R\ln\frac{V_2}{V_1} + \frac{\mu_0}{2C}\left(M_2^2 - M_1^2\right) \tag{20}$$

Then with the eq. (14) and converting to the exponential form

$$\frac{V_2}{V_1} = \left(\frac{T_1}{T_2}\right)^{\frac{3}{2}} e^{\frac{\mu_0 C}{2R}\left(\frac{H_2^2}{T_2^2} - \frac{H_1^2}{T_1^2}\right)} \tag{21}$$

And for a reversible adiabatic process in a paramagnetic gas we have

$$VT^{\frac{3}{2}}e^{-\frac{\mu_0 CH^2}{2RT^2}} = const \tag{22}$$

Considering now the Carnot cycle for a paramagnetic gas, in the first step, the reversible isothermal expansion (García-Colín Scherer, 1972)

$$dW = -PdV + \mu_0 HdM \tag{23}$$

For an isothermal process $dT=0$ and replacing (13), (14) in (23) and integrating

$$W_I = -RT_H \ln \frac{V_B}{V_A} + \frac{\mu_0 C}{2T_H}\left(H_B^2 - H_A^2\right) \tag{24}$$

In agreement with the first law of the thermodynamics, the absorbed heat in the first step is given by

$$Q_I = RT_H \ln \frac{V_B}{V_A} - \frac{\mu_0 C}{2T_H}\left(H_B^2 - H_A^2\right) \tag{25}$$

The second step is an adiabatic expansion where $Q_{II} = 0$ and the change in internal energy is equal to the work done $\Delta U_{II} = C_V(T_L - T_H)$

In the third step, the reversible isothermal compression

$$W_{III} = -Q_{III} = -RT_L \ln \frac{V_D}{V_C} + \frac{\mu_0 C}{2T_L}\left(H_D^2 - H_C^2\right) \tag{26}$$

For the final adiabatic compression again $Q_{IV} = 0$ and $\Delta U_{IV} = C_V(T_H - T_L)$, then as in the ideal gas the net work done in the two adiabatic steps is zero. In agreement with the first law (Dickerson, 1969; Nash, 1970), for a cyclic process $\Delta U = 0$ and the net work can be written as

$$W_{net} = W_I + W_{III} = -RT_H \ln \frac{V_B}{V_A} + \frac{\mu_0 C}{2T_H}\left(H_B^2 - H_A^2\right) - RT_L \ln \frac{V_D}{V_C} + \frac{\mu_0 C}{2T_L}\left(H_D^2 - H_C^2\right) \tag{27}$$

For the heat absorbed at $T_H$ is given by (25) and the efficiency is

$$\eta = \frac{-W_{net}}{Q_I} = \frac{RT_H \ln \frac{V_B}{V_A} - \frac{\mu_0 C}{2T_H}(H_B^2 - H_A^2) + RT_L \ln \frac{V_D}{V_C} - \frac{\mu_0 C}{2T_L}(H_D^2 - H_C^2)}{RT_H \ln \frac{V_B}{V_A} - \frac{\mu_0 C}{2T_H}(H_B^2 - H_A^2)} \quad (28)$$

and for the eq. (28) we have

$$\eta = \frac{-W_{net}}{Q_I} = 1 - \frac{RT_L \ln \frac{V_C}{V_D} + \frac{\mu_0 C}{2T_L}(H_D^2 - H_C^2)}{RT_H \ln \frac{V_B}{V_A} - \frac{\mu_0 C}{2T_H}(H_B^2 - H_A^2)} \quad (29)$$

From of (22) we can obtain

$$\frac{V_C}{V_D} = \frac{V_B}{V_A} e^{\left[\frac{\mu_0 C}{2RT_H^2}(H_A^2 - H_B^2) - \frac{\mu_0 C}{2RT_L^2}(H_D^2 - H_C^2)\right]} \quad (30)$$

Substituting eq. (30) in eq. (29)

$$\eta = \frac{-W_{net}}{Q_I} = 1 - \frac{RT_L \ln \frac{V_B}{V_A} - \frac{\mu_0 C T_L}{2T_H^2}(H_B^2 - H_A^2)}{RT_H \ln \frac{V_B}{V_A} - \frac{\mu_0 C}{2T_H}(H_B^2 - H_A^2)} \quad (31)$$

Rearranging (31)

$$\eta = \frac{-W_{net}}{Q_I} = 1 - \frac{RT_L \ln \frac{V_B}{V_A}\left[1 - \frac{\mu_0 C}{2RT_H^2}(H_B^2 - H_A^2)\right]}{RT_H \ln \frac{V_B}{V_A}\left[1 - \frac{\mu_0 C}{2RT_H^2}(H_B^2 - H_A^2)\right]} \quad (32)$$

And thus the eq.(32) takes the form

$$\eta = 1 - \frac{T_L}{T_H} \quad (33)$$

The efficiency of paramagnetic gas depends on the limits of maximum and minimal temperature of the cycle as in the ideal gas, photon gas and variable Chaplygin gas.

## 5. Conclusions

We have deduced an expression for the efficiency of a Carnot heat engine with an ideal paramagnetic gas, which is a function of the maximum and minimal temperature of the thermodynamic cycle. The study of paramagnetic gas can enrich the courses of thermodynamics, which contributes to a better compression of the thermal phenomena.

The thermodynamic equations that describe the paramagnetic gas are tractable mathematically and offer a wide comprehension of behavior in magnetic materials and about the physics of solid state. The study of ideal paramagnetic gas also serves as good introduction for further treatments in statistical and thermal physics of magnetic systems.

We have shown that the efficiency of a Carnot cycle with an paramagnetic gas is the same as that of a classical Carnot cycle as in the ideal gas, the photon gas and Chaplygin gas ( Dickerson, 1969; Leff,2002; Malaver, 2018). For a Carnot heat engine with a paramagnetic gas as working substance to be 100% efficient, temperature of hot reservoir $T_H$ must be infinite or zero for the cold reservoir $T_L$ and the second law says that a process cannot be 100% efficient in converting heat into work.

**References**

Ahmed, S.A.E., Babker. N.A.M., Fadel, M.T. (2019), A study on classes of magnetism, International Journal of Innovative Science, Engineering & Technology, Vol. 6, No.4, pp. 25-33.

Bender, C.M., Brody, D.C., Meister, B.K. (2000), Quantum mechanical Carnot engine, Journal of Physics A: Mathematical and General, Vol.33, No.24, 4427.

Chikazumi. S (2009). Physics of ferromagnetism, 2nd edition, Oxford: Oxford University Press, ISBN 9780199564811.

Dickerson, R. (1969). Molecular Thermodynamics, W.A.Benjamin, Inc, Menlo Park, California, ISBN: 0-8053-2363-5.

Eisberg, R., Resnick, R. (1991). Física Cuántica, Editorial Limusa, Séptima Edición, ISBN 968- 18-0419-8.

García-Colín Scherer, L. (1972). Introducción a la termodinámica clásica, Editorial Trillas, Segunda Edición.

Jackson, R. (2014), John Tyndall and the Early History of Diamagnetism, Annals of Science, Vol. 72, No.4, pp. 435–489.

Lee, L. (2001), Carnot cycle for photon gas?, Am.J.Phys, Vol. 69, pp. 874-878.

Leff, H.S. (2002), Teaching the photon gas in introductory physics, Am.J.Phys, Vol. 70, pp.792-797.

Malaver, M. (2015), Carnot engine model in a Chaplygin gas, Research Journal of Modeling and Simulation, Vol.2, No.2, pp.42-47.

Malaver, M. (2018), Carnot Heat Engine with a Variable Generalized Chaplygin Gas, Intertnational Journal of Astronomy, Astrophysics and Space Science, Vol.5, No.4, pp.45-50.

Miessler, G. L., Tarr, D. A. (2010). Inorganic Chemistry, Prentice Hall, 3rd edition, ISBN 0-13-035471-6.

Nash, L. (1970). Elements of Classical and Statistical Thermodynamics, Addison-Wesley Publishing Company, Inc, Menlo Park, California.

Wark, K., Richards, D. (2001). Termodinámica, McGraw-Hill Interamericana, Sexta Edición, ISBN: 84-481- 2829-X.